\documentclass[twocolumn,showpacs,prb,superscriptaddress]{revtex4}

\usepackage{graphicx}

\begin{document}

\title{Zero-field splitting of Kondo resonances in a carbon nanotube quantum dot}

\author{J.\ Nyg{\aa}rd}
\altaffiliation {Permanent address: Niels Bohr Institute,
University of Copenhagen, Electronic address: nygard@nbi.dk}

\affiliation{Department of Physics, Harvard University, Cambridge MA 02138}

\affiliation{Niels Bohr Institute and Nanoscience Center,
University of Copenhagen, Universitetsparken 5, DK-2100
Copenhagen, Denmark}

\author{W.F.\ Koehl} \affiliation{Department of Physics, Harvard University, Cambridge
MA 02138}

\author{N.\ Mason} \affiliation{Department of Physics, Harvard University, Cambridge
MA 02138}

\author{L.\ DiCarlo} \affiliation{Department of Physics, Harvard University, Cambridge
MA 02138}

\author{C. M.\ Marcus}

\affiliation{Department of Physics, Harvard University, Cambridge
MA 02138}

\date{\today} 

\begin{abstract}

We present low-temperature electron transport measurements on a
single-wall carbon nanotube quantum dot exhibiting Kondo
resonances at low temperature. Contrary to the usual behavior for
the spin-$1/2$ Kondo effect we find that the temperature
dependence of the zero bias conductance is nonmonotonic. In
nonlinear transport measurements low-energy splittings of the
Kondo resonances are observed at zero magnetic field. We
suggest that these anomalies reflect interactions
between the nanotube and a magnetic (catalyst) particle. The
nanotube device may effectively act as a ferromagnetically
contacted Kondo dot.
\end{abstract}

\pacs{72.15.Qm,73.22.-f,73.23.-b,73.63.Fg,73.63.Kv}

\maketitle

Transport measurements on individual single-wall carbon nanotubes have demonstrated their potential for nanoscale
electronics,\cite{mceuen02} ranging from high performance field-effect transistors to ideal one-dimensional quantum dots
with well-defined spin structure.\cite{cobden98,herrero04} The latter devices may even have prospects for solid-state
quantum computing based on the electronic spin,\cite{loss98} although fundamental issues regarding spin relaxation and
decoherence still need further investigations in addition to the requisite progress in device processing.

A prominent example of quantum coherence and spin physics in
quantum dots is the Kondo effect \cite{goldhaber98,kouwenhoven01}.
Below the Kondo temperature, $T_K$, an extended many-body state
arises from the interactions between a localized electron spin on
the dot and the conduction electrons in the leads. Kondo effects
have been found in quantum dots based on carbon
nanotubes\cite{nygard00} and in single molecule
transistors.\cite{park02} In contrast to semiconductor dots, these
systems allow for studies of Kondo effects in devices with
different types of contacts, e.g.\ superconducting or magnetic
electrodes. A nanotube Kondo dot coupled to superconducting
electrodes has been realised,\cite{buitelaar02ii} and magnetically
contacted quantum dots in the Kondo regime was reported recently
in molecular transistors.\cite{pasupathy04} We present here
experimental data on a nanotube quantum dot, where the Kondo
resonances are split at zero field. We interpret this as evidence
for coupling of the dot to a ferromagnetic impurity (in the form
of a catalyst nanoparticle).

\begin{figure}
\includegraphics[width=8.4cm]{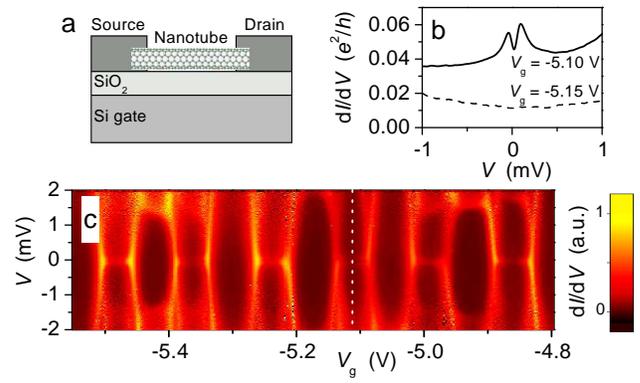}%
\caption{\label{fig-dIdV} (a) Schematic of nanotube device,
comprising a single-wall carbon nanotube, two Cr/Au electrodes
(source and drain), and a doped Si gate underneath the SiO$_2$ cap
layer. (b)  Differential conductance ${\rm d}I/{\rm d}V$ as a
function of source-drain voltage $V$ for gate voltages
$V_g=-5.10$~V (solid) and $V_g=-5.15$~V (dashed) at temperature
$T_{\rm el}=80$~mK. (c) Plot of ${\rm d}I/{\rm d}V$ as a function of
$V$ and $V_g$ (color online). The dashed line corresponds to the solid trace in
(b).}
\end{figure}

Our devices incorporate single-wall carbon nanotubes grown by
chemical vapor deposition (CVD) on an SiO$_2$
substrate.\cite{hafner01} Ferric iron nitrate nanoparticles
deposited from a solution in isopropyl alcohol acted as catalyst
for the CVD process, which was carried out in a tube furnace by
flowing methane  and hydrogen over the sample at 900$^\circ$C.
This process yielded mostly individual nanotubes with diameters in
the range 1-3~nm as determined by atomic force microscopy. The
nanotubes were contacted by thermally evaporated metal electrodes
(35~nm Au on 4~nm Cr), spaced by 250~nm and patterned by electron
beam lithography. Highly doped silicon below the 400 nm SiO$_2$
cap layer acted as a gate electrode, see Fig.~\ref{fig-dIdV}(a).
Electron transport measurements were carried out in a dilution
refrigerator with a base electron temperature of $T_{\rm el}\sim
80$~mK as estimated from the device characteristics. The
two-terminal conductance was measured using standard lock-in
techniques with $\sim 5$~mV ac excitation and voltage bias $V$
applied to the source with the drain grounded through a
low-impedance current amplifier.

In the appropriate range of back-gate voltage,  $V_g$, the room
temperature conductance is around 1.8~$e^2/h$ and only weakly
dependent on gate voltage $V_g$ (not shown), indicating that the
conducting nanotube is metallic. At low temperature, the
differential conductance, ${\rm d}I/{\rm d}V$, of the tube in the
same range of back-gate voltage shows a strong dependence on
$V_g$, as seen in Fig.~\ref{fig-dIdV}(b). The overall
characteristics of the device are more clearly seen from the 2D
plot of ${\rm d}I/{\rm d}V$ as a function of source-drain voltage
$V$ and gate voltage $V_g$ in Fig.~\ref{fig-dIdV}(c). The dominant
dark regions of low conductance are caused by Coulomb blockade
(CB) while the sloping bright lines are edges of the CB diamonds,
where the blockade is overcome by the finite source-drain
bias.\cite{kouwenhoven97}

Moreover, faint light horizontal ridges of high conductance around
zero bias are seen in Fig.~\ref{fig-dIdV}(c). These ridges
occur in a alternating manner, for every second electron added to
the nanotube dot. They are consistent with Kondo resonances
induced by the finite electron spin $S=1/2$ existing for an odd
number $N$ of electrons where an unpaired electron is localized
on the tube.\cite{nygard00} The zero bias resonances are absent
for the other regions (with even $N$) where the ground state spin
is $S=0$. A finite source-drain bias $eV\sim k T_K$ suppresses the
Kondo effect and the resonances are thus expected to appear as
thin horizontal lines at $V=0$ in Fig.~\ref{fig-dIdV}(c), or
equivalently as narrow peaks in ${\rm d}I/{\rm d}V$ as a function
of $V$. However, we note in Fig.~\ref{fig-dIdV}(b) that in fact a
fine splitting of the Kondo peak is observed.

\begin{figure}

\includegraphics[width=8.1cm]{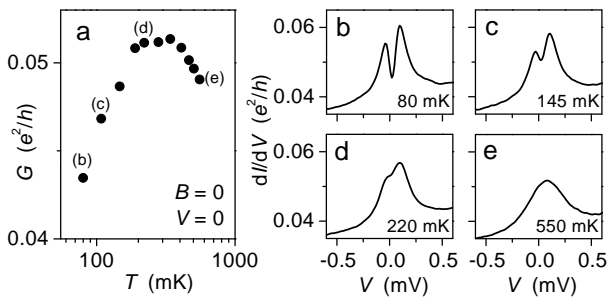}%

\caption{\label{fig-Tdep}(a) Temperature dependence of the
zero-bias conductance $G$ for the resonance shown in Fig.~1b
($V_g=-5.10$~V). (b)-(e) ${\rm d}I/{\rm d}V$ as a function of $V$
for the same resonance, at temperatures 80 (b), 145 (c), 220 (d),
and 550~mK (e).}
\end{figure}

Fig.~\ref{fig-Tdep} shows the temperature dependence of the zero bias
conductance. We find  a nonmonotonic temperature
dependence in contrast to the usual Kondo resonances where the linear
conductance, $G = {\rm d}I/{\rm d}V|_{V=0}$, scales as $G\sim
-\log(T)$ for $T\gg T_K$ and decreases monotonically with
increasing $T$. In the present device the conductance increases
with $T$ up to around 200~mK, where $G$ saturates at 0.05~e$^2$/h.
For higher $T$ (above $\sim 500$~mK) $G$ decreases as expected.
The accompanying ${\rm d}I/{\rm d}V$ plots in
Figs.~\ref{fig-Tdep}(b)-(d) show the peak profiles at various
temperatures. We note that the splitting is only observed at the
lowest temperatures, corresponding to the range where $G(T)$ is
increasing. Similar behavior is found for the other Kondo
resonances in Fig.~\ref{fig-dIdV}(c), although the magnitude of the
splittings vary between the different resonances which involve
different orbitals.

In a previous study of a semiconductor quantum dot in the Kondo
regime double resonance peaks were found for an $S=1$ 'two-stage'
Kondo effect, where two different energy scales result in the
appearance of a dip in the peak.\cite{vanderwiel02} However, this
scenario does not apply to  our case since the Kondo resonances
here only exist for $S=1/2$ as seen from the regular even-odd
alternations in Fig.~\ref{fig-dIdV}(c).

Application of an external magnetic field will normally cause the
$S=1/2$ Kondo resonance to split into two components with a peak
spacing $\Delta V= 2E_Z/e$, where  $E_Z= (1/2)g \mu_B B$ is the
electron Zeeman energy.\cite{goldhaber98} For the resonance in
Fig.~\ref{fig-dIdV}(b) the splitting is $\Delta V\approx 0.12$~mV,
which would correspond to an external magnetic field of 0.51~T,
since $g=2.0$ for the conduction electrons in metallic carbon
nanotubes.\cite{cobden98} Such a large external field offset
cannot exist in our experiment.\cite{magnetic}

\begin{figure}

\includegraphics[width=8.1cm]{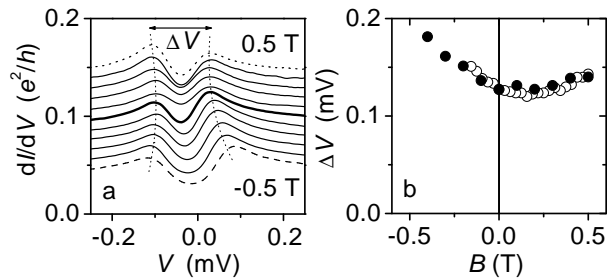}%

\caption{\label{fig-Bdep} (a) Magnetic field dependence of ${\rm
d}I/{\rm d}V$ as a function of source-drain voltage $V$  for the
peak in Fig.~2(b) at base $T_{\rm el}=80$~mK. The applied magnetic field $B$
was increased from $-0.5$~T (dashed) over 0~T (thick) to $0.5$~T
(dotted) in steps of 0.1~T. Subsequent curves are offset by
$0.01~e^2/h$ for clarity.  (b) Peak splitting $\Delta V$ as a
function of applied magnetic field $B$ for the data in (a) (solid)
and another series where $B$ was swept from 0.5~T to $-0.3$~T
(open).}
\end{figure}

The magnetic field dependence for the resonance in
Fig.~\ref{fig-Tdep}(b) is shown in Fig.~\ref{fig-Bdep}(a). The
resonance is split for all fields. At large field the splitting
grows as expected for the Zeeman splitting of a Kondo resonance.
However, the minimal splitting is in fact achieved at a small
finite field $B\sim 0.15$~mT, see Fig.~\ref{fig-Bdep}(b). The
splitting and asymmetry in field are significant and cannot be
explained based on a model of a purely non-magnetic quantum dot in
a weak external field.

Nanotubes can couple electrically to metal nanoparticles. For
example, Ref.~\cite{thelander01} demonstrates that a nonmagnetic
gold nanoparticle placed in the gap between two nanotube segments
can form a single-electron transistor with the nanotubes acting as
leads. The sensitivity of the electronic properties of nanotubes
to the presence of magnetic particles  adsorbed on the tube walls
has been proven by low-temperature STM measurements on tubes with
Co clusters.\cite{odom00,fiete02}  We interpret the splitting of
the Kondo peak as resulting from contact with a ferromagnetic
particle, presumably from the iron-containing catalyst material
used to grow the nanotubes. \cite{hafner01,stabledevice}
Theoretical work has shown the Kondo resonances persist when
coupling a quantum dot to two ferromagnetic leads, however,  the
Kondo peaks in ${\rm d}I/{\rm d}V$ may split, even in the absence
of an external magnetic field.\cite{martinek03,choi04} It is
anticipated that this result would also apply for dots coupled to
just one magnetic lead. It has recently been shown using numerical
renormalization group (NRG) theory
 that the conductance for such a
one-magnetic-lead quantum dot would exhibit a nonmonotonic
temperature dependence.\cite{sindel04} These results support our
interpretation that the splitting of the Kondo resonances in our
device reflects interactions of the quantum dot with a magnetic
impurity.





















In conclusion we have observed clear splittings of the Kondo
resonances in a nanotube dot at zero field. Interestingly, gaps below 1~meV were recently observed in another nanotube
quantum dot,\cite{babic04} but $T$ or $B$ dependences were not
reported, precluding detailed comparison to the present data. We
propose that the Kondo resonances in this study has probed the
effect of a magnetic impurity on electron transport in a carbon
nanotube device. Future experiments allowing gate-controlled
interaction with a ferromagnetic particle will provide important
further information on the effects of ferromagnetic particles\cite{oxide} on
quantum transport in nanotubes.
Until now all published studies on the effects of defects on
nanotube transport have considered nonmagnetic impurities or
defects, for instance charge traps in gate oxide, atomic defects,
contaminant particles, kinks, and normal metal particles.
Meanwhile, in most fabrication methods for single-wall nanotubes
magnetic particles from the catalyst remain in the material after
growth,\cite{hafner01,chen03} although a catalyst free route for
single-wall nanotube synthesis was devised
recently.\cite{derycke02} It should be feasible to probe nanotubes
coupled deliberately to individual magnetic particles, eg.\ by
moving them into contact by AFM manipulation.\cite{thelander01}
Likewise, it would  be desirable to achieve high transparency
contacts to lithographically defined ferromagnetic electrodes.
Nanotubes have already been contacted by magnetic
electrodes,\cite{tsukagoshi99} but in all reported studies the
transmission was too low to allow for Kondo resonances to form.
Notably, controlling or eliminating unwanted interactions with
magnetic materials will be crucial for experiments on spintronics
as well as  electron spin resonance and quantum computing in
nanotubes.

We acknowledge J.\ von Delft, J.\ Martinek, and M.\ Sindel for useful
discussions. The work was supported by ARO/ARDA (DAAD19-02-1-0039), NSF-NIRT (EIA-0210736), the Center for Imaging and
Mesoscale Structures at Harvard University, and the Danish
Technical Research Council STVF.


\begin{thebibliography}{99}



\bibitem{mceuen02} P.~L. McEuen, M.\ Fuhrer, and H.\ Park, IEEE Trans.\ on Nanotech.\ {\bf 1}, 78 (2002).



\bibitem{cobden98} D.~H.\ Cobden {\em et al.}, Phys.\ Rev.\ Lett.\ {\bf 81}, 681
  (1998).



\bibitem{herrero04} P.~Jarillo-Herrero {\em et al.}, Nature {\bf 429}, 389 (2004).



\bibitem{loss98} D.~Loss and D.P.~DiVincenzo, Phys.\ Rev.\ {\bf A 57}, 120 (1998).





\bibitem{goldhaber98} D.\ Goldhaber-Gordon {\em et al.}, Nature {\bf 391}, 156
(1998).



\bibitem{kouwenhoven01} L.~Kouwenhoven and L.~Glazman, Physics World {\bf 14} (1), 33 (2001).



\bibitem{nygard00} J.\ Nyg{\aa}rd, D.~H.\ Cobden, and P.~E.\ Lindelof, Nature {\bf 408}, 342 (2000).








\bibitem{park02} J.\ Park {\em et al.}, Nature {\bf 417}, 722 (2002),  W.~J.\ Liang {\em et al.}, Nature {\bf 417}, 725
(2002).



\bibitem{buitelaar02ii} M.\ Buitelaar {\em et al.}, Phys.\ Rev.\ Lett.\ {\bf 89}, 256801 (2002).

\bibitem{pasupathy04} A.~N.\ Pasupathy  {\em et al.}, Science {\bf 306}, 86 (2004).



\bibitem{hafner01} J.~H.\ Hafner {\em et al.}, J.\ Phys.\ Chem. B {\bf 105}, 743
(2001).



\bibitem{kouwenhoven97}See, e.g., L.~P.\ Kouwenhoven, C.~M.\ Marcus, P.~L.\ McEuen, S.\
  Tarucha, R.~M.\ Westervelt, and N.~S.\ Wingreen, in {\em Mesoscopic Electron
  Transport}, edited by L.~P.\ Kouwenhoven, G.\ Sch{\"o}n, and L.~L.\ Sohn
  (Kluwer, Dordrecht, The Netherlands, 1997).



\bibitem{vanderwiel02} W.G.\ van der Wiel {\em et al.}, Phys.\ Rev.\ Lett.\ {\bf 88}, 126803 (2002).



\bibitem{magnetic}
From the superconducting coil of our cryostat an offset field of
no more than 10~mT can be expected as shown in other experiments.
Moreover, care has been taken to avoid any magnetic material in
the setup (chip carrier, holder, leads, filters, cryostat parts
etc.).


\bibitem{thelander01} C.~Thelander {\em et al.}, Appl.\ Phys.\ Lett.\ {\bf 79}, 2106 (2001).



\bibitem{odom00} T.\ Odom {\em et al.}, Science {\bf 290}, 1549 (2000).



\bibitem{fiete02} G.~A.\ Fiete {\em et al.}, Phys.\ Rev.\ B {\bf 66}, 24431 (2002).



\bibitem{stabledevice} The overall diamond pattern of the dot
was stable for over a month while the device was kept at subkelvin
temperatures, i.e.\ the charge states did not change. Details in
the excited state spectrum and Kondo peak splittings changed
occasionally, possibly reflecting  varying interactions with an
impurity particle.



\bibitem{martinek03} J.~Martinek {\em et al.}, Phys.\ Rev.\ Lett.\ {\bf 91}, 127203 (2003), {\em ibid.\ } {\bf 91},
247202 (2003).

\bibitem{choi04} M.-S.\ Choi, D.\ Sanchez and R.\ Lopez, Phys.\ Rev.\ Lett.\ {\bf 92}, 056601
(2004).


\bibitem{sindel04} M.~Sindel and J.~von Delft (unpublished).



\bibitem{babic04} B.~Babic, T.~Kontos, and C.~Sch\"onenberger, condmat/0407193.



\bibitem{oxide} Even strong coupling to a ferromagnetic oxide
such as Fe$_3$O$_4$, which is half-metallic and believed to be
fully spin polarized, may influence the electronic properties like
for coupling to a magnetic metal.



\bibitem{chen03}F.\ Chen {\em et al.}, Appl.\ Phys.\ Lett.\ {\bf 83}, 4601 (2003).



\bibitem{derycke02} V.\ Derycke {\em et al.}, Nano Lett.\ {\bf 2}, 1046 (2002).




\bibitem{tsukagoshi99} K.\ Tsukagoshi, B.W.\ Alphenaar, and H.\ Ago, Nature {\bf 401}, 573 (1999).



\end{thebibliography}
\end{document}